# Design of Decoupling and Nonlinear PD Controller for Cruise Control of a Quadrotor

Hanum Arrosida
Department of Computer Control Engineering
Polytechnic of Madiun (PNM)
Madiun, Indonesia
hanumarrosida@pnm.ac.id

Rusdhianto Effendi
Department of Electrical Engineering
Institut Teknologi Sepuluh Nopember (ITS)
Surabaya, Indonesia
rusdhi@elect-eng.its.ac.id

Trihastuti Agustinah
Department of Electrical Engineering
Institut Teknologi Sepuluh Nopember (ITS)
Surabaya, Indonesia
trihastuti@ieee.org

Josaphat Pramudijanto
Department of Electrical Engineering
Institut Teknologi Sepuluh Nopember (ITS)
Surabaya, Indonesia
pramudijanto@gmail.com

*Abstract*— **Quadrotor is often used to accomplish various missions related to surveillance, territory mapping, search and rescue, and other purposes. Quadrotor is a nonlinear system with multiple input multiple output and has stability issue due to external disturbance. These characteristics lead to difficulty in cruise control of quadrotor automatically. Decoupling method is used to eliminate the interaction of other control on rotational motion, then the roll, pitch, and yaw angle can be controlled independently. Nonlinear PD controller is obtained from invers model of control signal on a quadrotor and it is used to control the translational motion in $x$ and $y$ axis with nonlinear dynamics because of the influence the rotational angle. Simulation results show that the proposed method can eliminate the control interaction of roll, pitch and yaw angle, hence it works like single input single output system and translational motion on $x$ and $y$ axis can achieve the expected trajectory precisely.**

## I. INTRODUCTION

Unmanned Aerial Vehicle (UAV) is applied in various missions related to surveillance [1]. Quadrotor is difficult to control automatically, because of high maneuverability, high nonlinearity, and under-actuated system characteristics [2]. Quadrotor must has a good stability while flying, especially in rotational and translational motion [3]. The mechanism of rotational motion and translational motion on quadrotor have control inputs that interact with each other and high nonlinearity. The interaction problem between the control inputs cause system instability. This can be overcome by dynamical decoupling technique which is able to make as if SISO system that can be controlled independently [4]. In the nonlinear dynamics of the system, nonlinear controller has better performance because it can approach the characteristics of the actual system. However, the controller design process carried out be very difficult because it requires a complete knowledge about the system. While the linear controller cannot compensate for the overall characteristics of the system because the linear controller is built based on the dynamic model that has been linearized [5]. In recent years, various control methods have been explored and applied to control the position and orientation of quadrotor. Conventional methods such as PID control is widely used to control the quadrotor motion, however nonlinearity effect cannot be compensated by the controller [6]. Nonlinear control method, such as the sliding mode controller is a nonlinear control method that can bring the state to the sliding surface. The state will be keep on the surface, even if external disturbance exist. However, there is disadvantage of sliding mode controller that is chattering phenomenon that causes a lot of heat that arise in electronic circuit [7].

This paper proposes decoupling method and nonlinear PD controller for stability and cruise control of a quadrotor.

The paper is organized as follows: section 2 presents a complete dynamic model of quadrotor. Control system design are presented in section 3. Simulations result of the proposed method are presented in section 4. Section 5 present a conclusion of this paper.

## II. DYNAMIC MODEL OF QUADROTOR

Fig.1 show that $E$ ($O, X, Y, Z$) as the earth coordinates and $b$ ($o', x, y, z$) as the body coordinates.

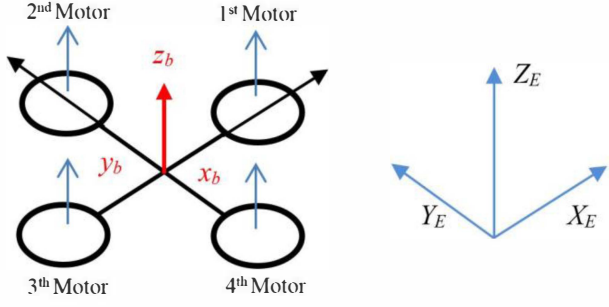

Fig. 1. Coordinate system of quadrotor

Based on the force that occurs in each motor, the torque of roll, pitch, and yaw motion can be calculated as shown on (1)-(4). Dynamic models of quadrotor are obtained through physical modeling based on Newton-II law for translational motion and Newton-Euler law for rotational motion as shown on (5)-(10).

$$U_1 = F_{T1} + F_{T2} + F_{T3} + F_{T4} \tag{1}$$

$$U_2 = F_{T2} - F_{T4} \tag{2}$$

$$U_3 = F_{T1} - F_{T3} \tag{3}$$

$$U_4 = d(F_{T1} + F_{T3} - F_{T2} - F_{T4}) \tag{4}$$

$$\ddot{x} = (\sin\phi \sin\psi + \cos\phi \sin\theta \cos\psi)\frac{U_1}{m} \tag{5}$$

$$\ddot{y} = (-\sin\phi \cos\psi + \cos\phi \sin\theta \sin\psi)\frac{U_1}{m} \tag{6}$$

$$\ddot{z} = \frac{U_1}{m}\cos\phi \cos\theta - g \tag{7}$$

$$\ddot{\phi} = \dot{p} = \frac{U_2 l}{J_{xx}} - \frac{qr}{J_{xx}}(J_{xx} - J_{yy}) \tag{8}$$

$$\ddot{\theta} = \dot{q} = \frac{U_3 l}{J_{yy}} - \frac{pr}{J_{yy}}(J_{xx} - J_{zz}) \tag{9}$$

$$\ddot{\psi} = \dot{r} = \frac{U_4}{J_{zz}} - \frac{pq}{J_{zz}}(J_{yy} - J_{xx}) \tag{10}$$

The values of $J_{xx}$ is inertia in $x$ axis, $J_{yy}$ is inertia in $y$ axis, and $J_{zz}$ is inertia in $z$ axis. Parameter values are shown in Table 1.

TABLE I. PARAMETER VALUES OF SYSTEM

| Parameter | Symbol | Value | Unit |
|---|---|---|---|
| PWM input from actuator | $K$ | 120 | N |
| Radius of *quadrotor* propeller | $L$ | 0.2 | M |
| Inertia moment on x axis | $J_{xx}$ | 0.03 | kg.m² |
| Inertia moment on y axis | $J_{yy}$ | 0.03 | kg.m² |
| Inertia moment on z axis | $J_{zz}$ | 0.04 | kg.m² |
| Mass of *quadrotor* | $m$ | 3.499 | kg |
| Actuator *Bandwidth* | $\omega$ | 15 | rad/sec |

Source: [8]

## III. CONTROL SYSTEM DESIGN

Quadrotor is a nonlinear system with multiple input multiple output and has stability issue due to external disturbance. This problem will be overcome by the nonlinear PD control method for controlling the translational motion and decoupling method with a PID controller to control the rotational motion, so quadrotor can cruise on the expected trajectory in stable condition. Control strategy of this system is shown in Fig 2.

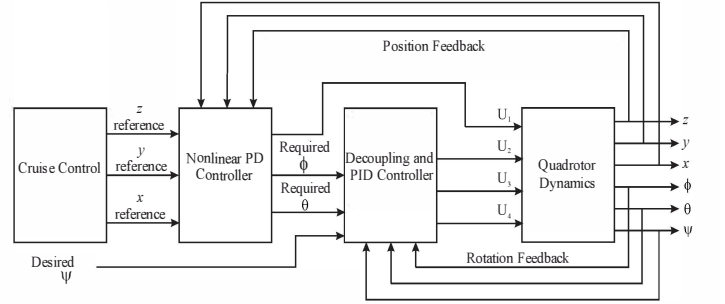

Fig. 2. Control system strategy

### A. Decoupling Mechanism of Rotational Angle

Based on the dynamic model of rotational motion is known that roll, pitch and yaw angle are interact. To eliminate the interaction controls of rotational angle is used decoupling method that can control the angle of roll, pitch and yaw independently. Simplification rotational angle model is defined $k_1 = (J_{yy} - J_{zz}) / J_{xx}$, $k_2 = l / J_{xx}$, $k_3 = (J_{zz} - J_{xx}) / J_{yy}$, $k_4 = l / J_{yy}$, $k_5 = (J_{xx} - J_{yy}) / J_{zz}$, $k_6 = 1 / J_{zz}$, can be obtained a new fashion for rotational motion of quadrotor are shown in (11). Decoupling mechanism shown in (12)-(15).

$$\begin{bmatrix} \ddot{\phi} \\ \ddot{\theta} \\ \ddot{\psi} \end{bmatrix} = \begin{bmatrix} k_1 qr + k_2 U_2 \\ k_3 pr + k_4 U_3 \\ k_5 pq + k_6 U_4 \end{bmatrix} \tag{11}$$

$$\begin{bmatrix} \dot{\phi} \\ \dot{p} \\ \dot{\theta} \\ \dot{q} \\ \dot{\psi} \\ \dot{r} \end{bmatrix} = \begin{bmatrix} p \\ k_1 qr \\ q \\ k_3 pr \\ r \\ k_5 pq \end{bmatrix} + \begin{bmatrix} 0 \\ k_2 U_2 \\ 0 \\ k_4 U_3 \\ 0 \\ k_6 U_4 \end{bmatrix} \tag{12}$$

$$\begin{aligned} U_2^* &= k_1 qr + k_2 U_2 \\ U_2 &= -\frac{1}{k_2}(k_1 qr) + \frac{U_2^*}{k_2} \\ U_3^* &= k_3 pr + k_4 U_3 \\ U_3 &= -\frac{1}{k_4}(k_3 pr) + \frac{U_3^*}{k_4} \\ U_4^* &= k_5 pq + k_6 U_4 \\ U_4 &= -\frac{1}{k_6}(k_5 pq) + \frac{U_4^*}{k_6} \end{aligned} \tag{13}$$

$$\begin{bmatrix} \dot\phi \\ \dot p \\ \dot\theta \\ \dot q \\ \dot\psi \\ \dot r \end{bmatrix} = \begin{bmatrix} 0 & 1 & 0 & 0 & 0 & 0 \\ 0 & 0 & 0 & 0 & 0 & 0 \\ 0 & 0 & 0 & 1 & 0 & 0 \\ 0 & 0 & 0 & 0 & 0 & 0 \\ 0 & 0 & 0 & 0 & 0 & 1 \\ 0 & 0 & 0 & 0 & 0 & 0 \end{bmatrix} \begin{bmatrix} \phi \\ p \\ \theta \\ q \\ \psi \\ r \end{bmatrix} + \begin{bmatrix} 0 \\ k_1 qr \\ 0 \\ k_3 pr \\ 0 \\ k_5 pq \end{bmatrix} + \begin{bmatrix} 0 \\ k_2\left(-\frac{1}{k_2}(k_1 qr) + \frac{U_2^*}{k_2}\right) \\ 0 \\ k_4\left(-\frac{1}{k_4}(k_3 pr) + \frac{U_3^*}{k_4}\right) \\ 0 \\ k_6\left(-\frac{1}{k_6}(k_5 pq) + \frac{U_4^*}{k_6}\right) \end{bmatrix} \quad (14)$$

$$\begin{bmatrix} \dot\phi \\ \dot p \\ \dot\theta \\ \dot q \\ \dot\psi \\ \dot r \end{bmatrix} = \begin{bmatrix} 0 & 1 & 0 & 0 & 0 & 0 \\ 0 & 0 & 0 & 0 & 0 & 0 \\ 0 & 0 & 0 & 1 & 0 & 0 \\ 0 & 0 & 0 & 0 & 0 & 0 \\ 0 & 0 & 0 & 0 & 0 & 1 \\ 0 & 0 & 0 & 0 & 0 & 0 \end{bmatrix} \begin{bmatrix} \phi \\ p \\ \theta \\ q \\ \psi \\ r \end{bmatrix} + \begin{bmatrix} 0 \\ U_2^* \\ 0 \\ U_3^* \\ 0 \\ U_4^* \end{bmatrix} \quad (15)$$

### B. Control Design of Nonlinear PD for Translational Motion

System dynamics of translational motion in the $x$ axis and the $y$ axis are nonlinear fashion because of the influence rotational angle, thereby translational motion in the $x$ axis and $y$ axis are controlled by nonlinear PD controller. Structure of nonlinear PD controller is obtained from invers model of control signal on a quadrotor, it is described on Fig 3.

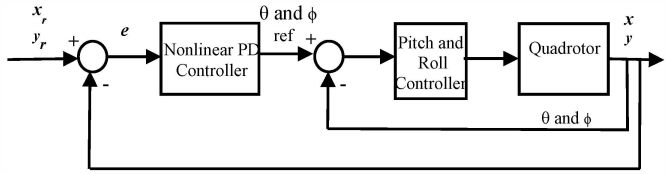

Fig. 3. Nonlinear PD Controller Block Diagram

Nonlinear PD controller of translational motion on $x$ axis can be obtained based on nonlinear dynamic of translational motion on $x$ axis and formulation of linear PD controller. Determination of nonlinear controller is expressed in (16).

$$\ddot x = (\sin\phi\sin\psi + \cos\phi\sin\theta\cos\psi)\frac{U_1}{m}$$
$$u_x = (\sin\phi\sin\psi + \cos\phi\sin\theta\cos\psi)\frac{U_1}{m}$$
$$\ddot x = u_x$$
$$e = x - x_r$$
$$u_x = K_p(\tau_d s + 1)(x - x_r)$$
$$(\sin\phi\sin\psi + \cos\phi\sin\theta\cos\psi)\frac{U_1}{m} = K_p(\tau_d s + 1)(x - x_r)$$
$$\sin\theta = \frac{K_p(\tau_d s + 1)(x - x_r) - \sin\phi\sin\psi\frac{U_1}{m}}{\cos\phi\cos\psi\frac{U_1}{m}}$$
$$\theta = \sin^{-1}\left(\frac{K_p(\tau_d s + 1)(x - x_r) - \sin\phi\sin\psi\frac{U_1}{m}}{\cos\phi\cos\psi\frac{U_1}{m}}\right) \quad (16)$$

The value of gain $K_p$ and $K_d$ are obtained from transfer function of linear PD controller shown in (17).

$$\frac{x}{x_r} = \frac{\frac{K_p(\tau_d s + 1)}{s^2}}{1 + \frac{K_p(\tau_d s + 1)}{s^2}} = \frac{K_p(\tau_d s + 1)}{s^2 + K_p(\tau_d s + 1)} \quad (17)$$

$$\frac{x}{x_r} = \frac{\tau_d s + 1}{\frac{1}{K_p}s^2 + \tau_d s + 1}$$

Pole placement are determined in 0.7 and 4.3, then the value of $K_p$ and $K_d$ gain can be obtained in (18).

$$\frac{x}{x_r} = \frac{a}{s+4.3} + \frac{b}{s+0.7} = \frac{as + 0.7a + bs + 4.3b}{s^2 + 5s + 3.01}$$

$$\frac{x}{x_r} = \frac{\frac{1}{3.01}(a+b)s + \frac{1}{3.01}(0.7a + 4.3b)}{\frac{1}{3.01}s^2 + \frac{5}{3.01}s + 1} = \frac{\tau_d s + 1}{\frac{1}{K_p}s^2 + \tau_d s + 1} \quad (18)$$

$$K_p = 3.01$$
$$\tau_d = \frac{5}{3.01} = 1.66$$
$$K_d = 5.0$$

Nonlinear PD controller of translational motion on $y$ axis is described on (19).

$$\ddot y = (-\sin\phi\cos\psi + \cos\phi\sin\theta\sin\psi)\frac{U_1}{m}$$
$$u_y = (-\sin\phi\cos\psi + \cos\phi\sin\theta\sin\psi)\frac{U_1}{m}$$
$$\ddot y = u_y$$
$$((\frac{U_1}{m}\cos\psi)^2 - (\frac{U_1}{m}\sin\theta\sin\psi)^2)\sin^2\phi + (\frac{U_1}{m}\sin\theta\sin\psi)^2 = K_p(\tau_d s + 1)(y - y_r)$$
$$((\frac{U_1}{m}\cos\psi)^2 - (\frac{U_1}{m}\sin\theta\sin\psi)^2)\sin^2\phi = K_p(\tau_d s + 1)(y - y_r) - (\frac{U_1}{m}\sin\theta\sin\psi)^2$$
$$\sin^2\phi = \frac{K_p(\tau_d s + 1)(y - y_r) - (\frac{U_1}{m}\sin\theta\sin\psi)^2}{((\frac{U_1}{m}\cos\psi)^2 - (\frac{U_1}{m}\sin\theta\sin\psi)^2)}$$
$$\sin\phi = \sqrt{\frac{K_p(\tau_d s + 1)(y - y_r) - (\frac{U_1}{m}\sin\theta\sin\psi)^2}{((\frac{U_1}{m}\cos\psi)^2 - (\frac{U_1}{m}\sin\theta\sin\psi)^2)}} \quad (19)$$
$$\phi = \sin^{-1}\left(\sqrt{\frac{K_p(\tau_d s + 1)(y - y_r) - (\frac{U_1}{m}\sin\theta\sin\psi)^2}{((\frac{U_1}{m}\cos\psi)^2 - (\frac{U_1}{m}\sin\theta\sin\psi)^2)}}\right)$$

## IV. RESULT AND ANALYSIS

Cruise control simulation perform on circle trajectory is used to accomplish as search and rescue mission. Expected trajectory that has to followed by quadrotor shown in Fig.4.

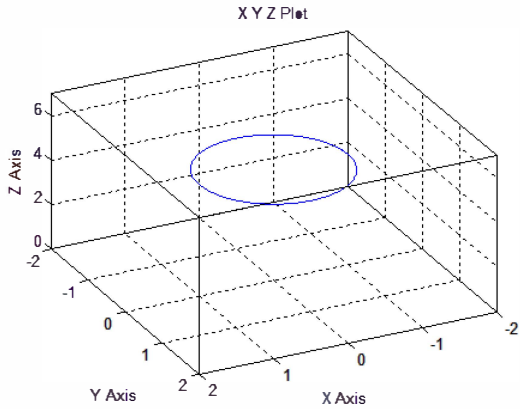

Fig. 4. Circle trajectory of cruise motion

In this discussion will be compared between cruise control using nonlinear PD controller based on invers model of control signal on a quadrotor and cruise control using nonlinear PD controller based on Gonzales's paper [5]. Actual responses of cruise control are shown in Fig.5.

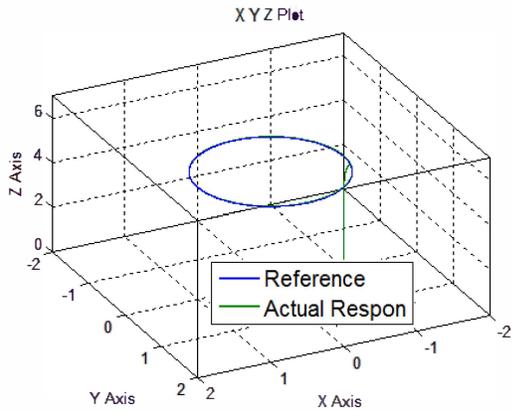

(a)

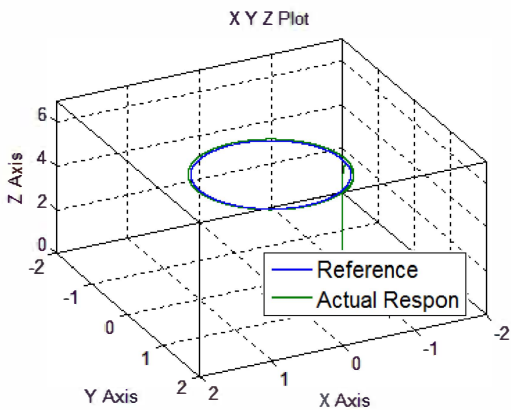

(b)

Fig. 5. (a) Cruise control using nonlinear PD controller based on invers model
(b) Cruise control using nonlinear PD controller based on paper [5]

Simulation of cruise control to accomplish territory mapping perform on square trajectory. The responses of cruise control on square trajectory shown in Fig.6.

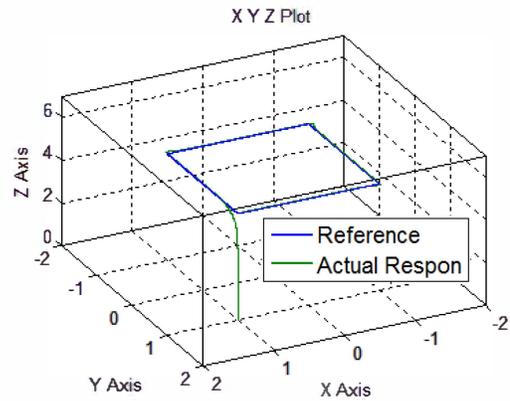

(a)

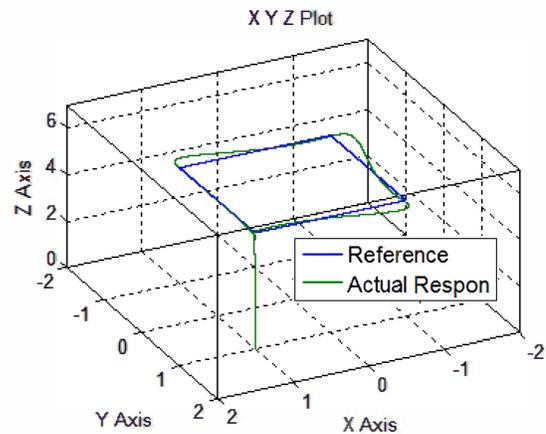

(b)

Fig. 6. (a) Cruise control on square trajectory using nonlinear PD controller based on invers model

(b) Cruise control on square trajectory using nonlinear PD controller based on paper [5]

Based on simulation result, it can be concluded that using of nonlinear PD controller based on invers model is better than nonlinear PD controller based on paper [5] because quadrotor can follow the expected trajectory precisely with rise time 1 second, settling time 2 seconds, and overshoot 4%. When using nonlinear PD controller based on Gonzales's paper [5] have rise time 6 seconds, settling time 12 seconds and overshoot 11%.

## V. CONCLUSION

Decoupling method is applied to the roll, pitch, and yaw angle to make a system with multiple input multiple output works as a single input single output system by eliminate the interaction between the control signal, hence the rotational angle can be controlled independently and stability of system during cruise control can be achieved.

The use of nonlinear PD controller based on invers model of signal control in translational motion *x* and *y* axis is better than nonlinear PD controller based on paper because system takes 2 seconds to achieve settle condition with overshoot 4%, after that can follow the expected trajectory precisely, compared using the nonlinear PD controller based on Gonzales's paper have settling time 12 second and overshoot 11%.